\documentclass[twocolumn]{aa}
\usepackage[colorlinks=true]{hyperref}
\usepackage{graphicx}
\usepackage{rotating}
\usepackage{longtable}
\usepackage{pdflscape}
\usepackage{comment}
\usepackage{threeparttable}
\usepackage{txfonts}
\usepackage{tabularx}
\usepackage{amsmath}

\begin{document}

%\title {The spin measurement of the black hole in GRS 1716--249 constrained with the continuum and X-ray reflection component}
%\title {A low-spin black hole in GRS 1716--249: constrained with continuum and reflection components under new distance}
\title {A Revised Spin of the Black Hole in GRS 1716-249 with a New Distance}

\titlerunning{The spin of GRS 1716--249}

\author
{
S. J. Zhao\inst{1,2} 
\and L. Tao\inst{1}\thanks{E-mail: taolian@ihep.ac.cn}
\and Q. Q. Yin\inst{1}\thanks{E-mail:yinqq@ihep.ac.cn}
\and S. N. Zhang\inst{1,2}
\and R. C. Ma\inst{1,2}
\and P. P. Li\inst{1,2}
\and Q. C. Zhao\inst{1,2}
\and M. Y. Ge\inst{1}
\and L. Zhang\inst{1}
\and J. L. Qu\inst{1}
\and S. Zhang\inst{1}
\and X. Ma\inst{1}
\and Y. Huang\inst{1}
\and J. Q. Peng\inst{1,2}
\and Y. X. Xiao\inst{1,2}
}
% List of institutions

\institute{Key Laboratory of Particle Astrophysics, Institute of High Energy Physics, Chinese Academy of Sciences, Beijing 100049, China
\and University of Chinese Academy of Sciences, Chinese Academy of Sciences, Beijing 100049, China
 }

\authorrunning{S.J.Zhao et al.}
\date{Received ; accepted }

% \abstract{}{}{}{}{} 
% 5 {} token are mandatory
 
\abstract 
{GRS 1716--249 is a stellar-mass black hole in a low-mass X-ray binary that underwent a gaint outburst in 2016/17. In this paper we use simultaneous observations of Insight-HXMT and NuSTAR to determine its basic parameters. The observations were performed during the softest part of the outburst, and the spectra show clear thermal disk emission and reflection features. We have fitted the X-ray energy spectra using the joint fitting method of the continuum and reflection components with the {\tt kerrbb2} + {\tt relxill} model. Since there is a possibility that the distance to this source was previously underestimated, we use the latest distance parameter of 6.9\,kpc in our study, in contrast to previous work in which the distance was set at 2.4\,kpc. Through spectral fitting of fixing black hole mass at 6.4\,$M_{\rm \odot}$, we observe a strong dependence of the derived spin on the distance: $a_{*}=0.972_{-0.005}^{+0.004}$ at an assumed distance of 2.4\,kpc and $a_{*}=0.464_{-0.007}^{+0.016}$ at an assumed distance of 6.9\,kpc, at a confidence level of 90\%. If considering the uncertainties in the distance and black hole mass, there will be a wider range of spin with $a_{*}$ < 0.78. The fitting results with the new distance indicate that GRS 1716--249 harbors a moderate spin black hole with an inclined ($i\sim 40-50^{\circ}$) accretion disk around it. Additionally, we have also found that solely using the method of the reflection component fitting but ignoring the constraints on the spin from the accretion disk component will result in an extremely high spin.}

\keywords{X-ray binaries --- black hole physics --- accretion, accretion disk --- stars: individual: (GRS 1716--249)}

\maketitle

%________________________________________________________________

\section{Introduction}
Spin is an important parameter of a black hole (BH), which, together with its mass, determines the spacetime around it and influences the accretion of matter and the formation of jets (for a review, see \citealt{2021Reynolds}). Generally, the BH spin $a_{*}$ is defined as $a_{*}=Jc/GM^{2}$, where $J$ represents the angular momentum of the black hole, $G$ stands for the gravitational constant and $c$ denotes the speed of light. The magnitude of the spin determines the radius of the innermost stable circular orbit (ISCO), $R_{\rm ISCO}$. A larger value of $a_{*}$ corresponds to a smaller $R_{\rm ISCO}$. For example, when $a_{*}=0$ (Schwarzschild black hole), $R_{\rm ISCO}$ is equal to 6\,$R_{\rm g}$ (where $R_{\rm g}=GM/c^{2}$ is called the gravitational radius); when $a_{*}=1$ (extreme Kerr black hole), $R_{\rm ISCO}$ is equal to 1\,$R_{\rm g}$. Currently, there are two main methods for measuring black hole spin through spectral fitting: continuum fitting \citep{1997Zhang,2014McClintock} and reflection component fitting \citep{1989Fabian,2008Reynolds,2014Garcia,2018Garcia}. Both methods require the common prerequisite that the inner edge of the accretion disk is located at the position of the ISCO ($R_{\rm in} = R_{\rm ISCO}$).

%In the case of low accretion rates (<30\% of the Eddington limit), $R_{\rm ISCO}$ represents the minimum possible value for the inner radius ($R_{\rm in}$) of the accretion disk. When $R_{\rm in} > R_{\rm ISCO}$, we refer to the accretion disk as truncated.

%The development and application of the continuum fitting method primarily rely on   . Within the framework of the standard accretion disk theory,

According to the standard accretion disk theory \citep{1973Shakura,1973Novikov}, the spin of a black hole affects the temperature of the inner disk, which in turn affects the thermal emission of the accretion disk. The method of continuum fitting is precisely used to constrain the spin of the black hole by fitting the spectrum of the disk. However, the use of this method requires that the distance to the source ($D$), the inclination of the accretion disk ($i$), and the mass of the black hole ($M_{\rm BH}$) are known (the reasons can be found in \citealt{2011McClintock}). Two commonly used models for fitting the disk component are {\tt kerrbb} \citep{2005Li} and {\tt kerrbb2} \citep{2006McClintock_kerrbb2}, which consider the full relativistic effects. The key fitting parameters of these models include the black hole spin and the mass accretion rate ($\dot{M}$).

The typical reflection features are the iron line (at $\sim$6--7\,keV) and the Compton hump (at $\sim$20--30\,keV). The iron line emitted from the inner disk close to the black hole is influenced by the Doppler effect, relativistic beaming effect, and gravitational redshift, resulting in a broadened and distorted profile \citep{2003Reynolds,2007Miller}. The gravitational redshift mainly affects the red wing of the iron line. The larger the black hole spin, the further the red wing extends to lower energies. This is because a larger spin allows the inner disk to be closer to the black hole and experience a stronger gravitational potential well. The core of the reflection component fitting is to fit the profile of the iron line. This method does not depend on $M_{\rm BH}$ and $D$ and can also be used to constrain $i$.

In theory, there are billions of stellar-mass black holes in the Milky Way galaxy \citep{1994Brown}. However, there are only around twenty X-ray binary systems containing a dynamically confirmed black hole \citep{2006Remillard,2010Ozel,2016Corral-Santana}. GRS 1716--249 is one of them. This source was discovered in 1993 by the CGRO/BATSE and Granat/SIGMA telescopes \citep{1993Harmon,1993Ballet}. After about 23 years of quiescence, GRS 1716--249 had another outburst detected by MAXI on December 18, 2016 \citep{2016Negoro}. During this outburst, when the source was in the hard intermediate state, it is believed that the accretion disk was a standard disk with the inner edge located at the ISCO \citep{2019Bassi}. Additionally, the spectrum showed significant power-law (PL) components from the corona, as well as prominent reflection features (broad iron line and Compton hump) \citep{2019Bassi,2019Tao}. Therefore, this source presents an ideal case for measuring the black hole spin because both continuum fitting and reflection component fitting methods can be combined to simultaneously constrain the parameters of the black hole.

The system parameters of this source, especially the distance, have not been well constrained. \cite{1994della_Valle} estimated the distance to be 2.2--2.8\,kpc. \cite{1996Masetti} provided a lower limit on the black hole mass of 4.9\,$M_{\rm \odot}$, a companion star mass of $\sim$1.6\,$M_{\rm \odot}$, an orbital period of $\sim$14.7\,hr, and a distance of $2.4\pm0.4$\,kpc. By fixing the distance at 2.4\,kpc, \cite{2019Tao} found the black hole spin of $a_{*}>0.92$, the accretion disk inclination of $i \sim 40-50^{\rm \circ}$, and the black hole mass of $M_{\rm BH}$ < 8\,$M_{\rm \odot}$, and \cite{2021Chatterjee} used a two-component advective flow (TCAF) model to constrain the black hole mass to be between 4.5--5.9\,$M_{\rm \odot}$. However, \cite{2022Saikia} argue that the distance of 2.4\,kpc is an underestimate. They conservatively estimated the distance of GRS 1716--249 to be 4--17\,kpc based on the global optical/X-ray correlation and suggested a most likely value of $D$ $\sim$ 4--8\,kpc based on the dynamics of the binary system. \cite{2023Casares} presented evidence for a 6.7\,hr orbital period and used the empirical relationship between the quiescent $r$-band magnitude and the orbital period to constrain the distance to $6.9\pm1.1$\,kpc, consistent with the results given by \cite{2022Saikia}. Furthermore, \cite{2023Casares} also provided an orbital inclination of $61\pm15^{\circ}$ and a BH mass of $6.4_{-2.0}^{+3.2}$,$M_{\odot}$ at 68\% confidence.

Due to the distance-dependent nature of continuum fitting methods, different distances may yield different spin measurement results. Therefore, in this paper, we used two different distance parameters, the previously used 2.4\,kpc \citep{1996Masetti} and the most recent 6.9\,kpc provided by \cite{2023Casares}, to highlight the impact of distance changes on spin measurements. We employed a joint fitting method using continuum and reflection spectra to constrain the spin, utilizing data from Insight-HXMT and NuSTAR simultaneous observations during the hard intermediate state in the 2016--2017 outbrst. In the following section (Sect.~\ref{sec:obs}), we will describe the observations and data reduction. Sect.~\ref{sec:res} will present the spectral fitting and results. The obtained results will be discussed in Sect.~\ref{sec:dis} and summarized in Sect.~\ref{sec:con}.

%__________________________________________________________________

\section{Observations and data reduction}
\label{sec:obs}

\begin{figure}
    \centering
    \includegraphics[scale=0.32]{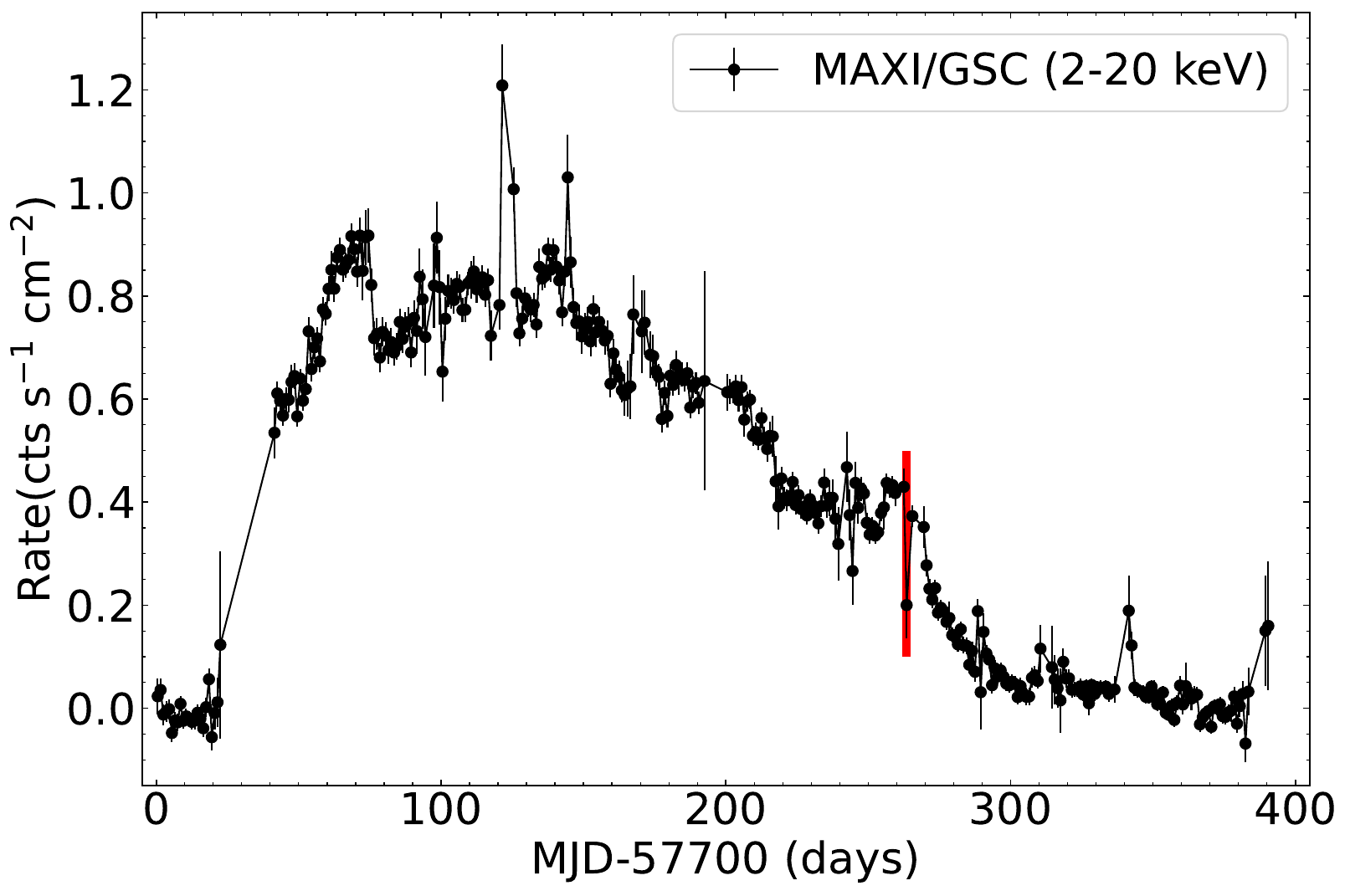}
    \caption{MAXI/GSC light-curve (2--20 keV) of GRS 1716--249. The solid red vertical line represents the simultaneous Insight-HXMT and NuSTAR observations used in this work.}
    \label{fig:lc}
\end{figure}

\begin{figure}
    \centering
    \includegraphics[scale=0.32]{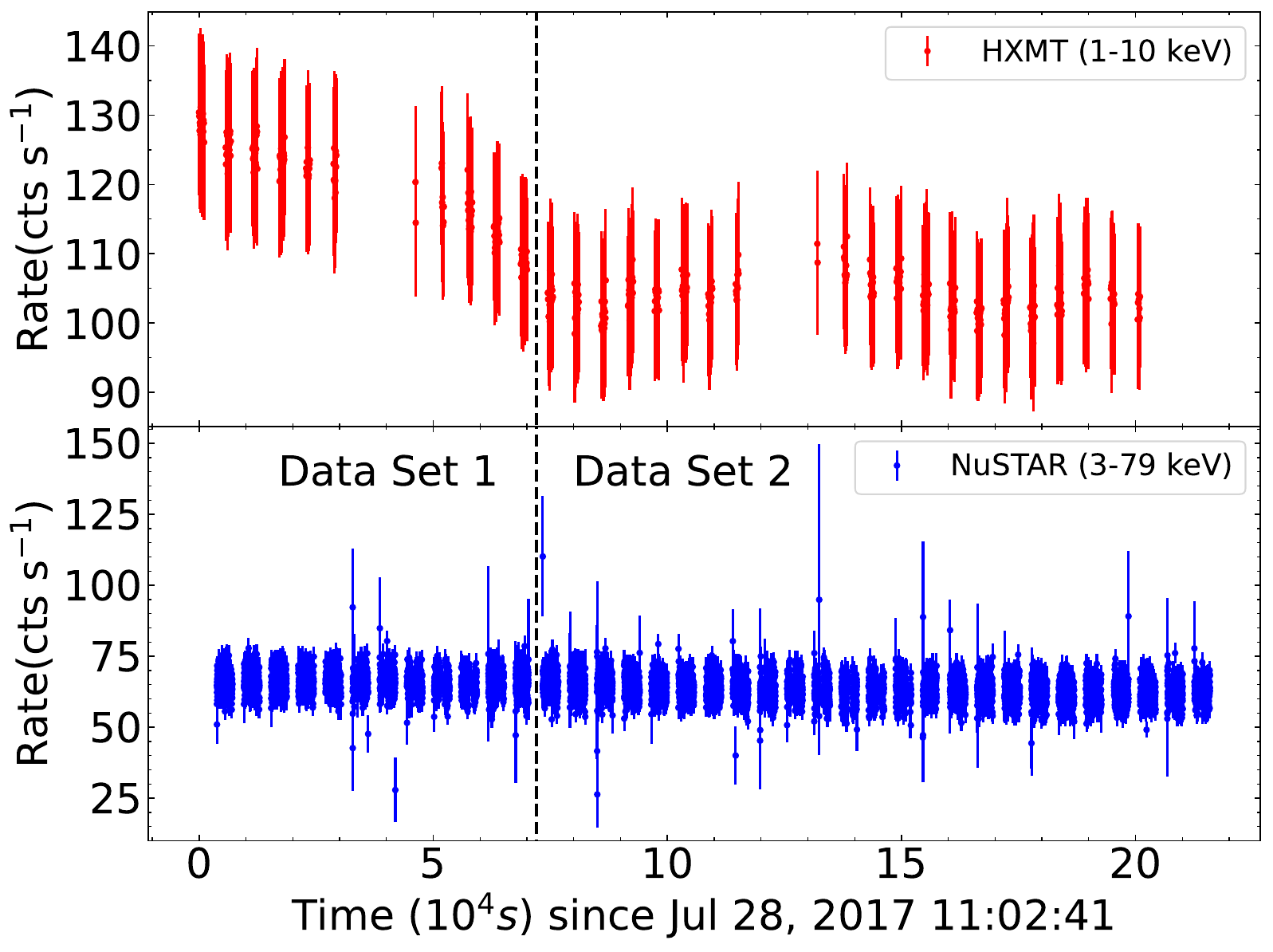}
    \caption{The light curves of Insight-HXMT/LE (top pannel) and  NuSTAR/FPMA (bottom panel). The black dashed vertical line represents the division point for the two data sets.}
    \label{fig:lc2}
\end{figure}
%%%%%%%%%%%%%%%%%%%%%%%%%%%%%%%%%%%%%%%%%%%%%%%%%%%%%%%%%%%%%%%
\begin{table*}
\centering
\renewcommand{\arraystretch}{1.5}
\caption{Observation information of Insight-HXMT and NuSTAR}
\label{tab:obs}
\begin{tabularx}{\textwidth}{@{}>{\centering\arraybackslash}X>{\centering\arraybackslash}X>{\centering\arraybackslash}X>{\centering\arraybackslash}X>{\centering\arraybackslash}X>{\centering\arraybackslash}X@{}}
\hline
\hline
Mission & ObsID &  Start Time &  End Time & Effective Exposure  & Net Count Rate \\
  &   &    &    & (s)  & (counts $\rm s^{-1}$) \\
\hline                                             
Insight-HXMT    & P0114335002 & 2017 Jul 28 11:02:41 & 2017 Jul 30 23:37:42 & 26102.6  &  92.6                   \\
NuSTAR  & 90301007002 & 2017 Jul 28 12:06:09 & 2017 Jul 30 23:21:09 & 89256.0    &   47.1                  \\
\hline
\end{tabularx}
\begin{tablenotes} 
\item Notes. For Insight-HXMT and NuSTAR, the observation logs of LE and FPMA are listed as a representation, respectively.
\end{tablenotes} 
\end{table*}

%%%%%%%%%%%%%%%%%%%
\begin{table*}
\centering
\renewcommand{\arraystretch}{1.5}
\caption{Information of Data Set 1 and 2.}
\label{tab:datasets}
\begin{tabularx}{\textwidth}{@{}>{\centering\arraybackslash}X>{\centering\arraybackslash}X>{\centering\arraybackslash}X>{\centering\arraybackslash}X>{\centering\arraybackslash}X@{}}
\hline
\hline
Mission  &  Start Time &  End Time & Effective Exposure  & Net Count Rate \\
   &    &    &  (s)  & (counts $\rm s^{-1}$) \\
\hline
\multicolumn{5}{c}{Data Set 1}                                              \\
Insight-HXMT     & 2017 Jul 28 11:02:41 & 2017 Jul 29 07:02:41 & 8981.5  & 101.4                        \\
NuSTAR   & 2017 Jul 28 12:06:09 & 2017 Jul 29 07:02:41 & 27729.6    &  48.4                      \\
\hline
\multicolumn{5}{c}{Data Set 2}                                                                         \\
Insight-HXMT     & 2017 Jul 29 07:02:41 &  2017 Jul 30 23:37:42 & 19691.7   & 84.3                     \\
NuSTAR   & 2017 Jul 29 07:02:41 & 2017 Jul 30 23:21:09 & 61532.9      & 46.6                   \\
 \hline

\end{tabularx}
\begin{tablenotes} 
\item Notes. For Insight-HXMT and NuSTAR, the observation logs of LE and FPMA are listed as a representation, respectively.
\end{tablenotes} 
\end{table*}
%%%%%%%%%%%%%%%%%%%

In contrast to \cite{2019Tao}, which utilized simultaneous Swift and NuSTAR data, in this study we use simultaneous Insight-HXMT and NuSTAR data. Insight-HXMT can provide high-statistics spectra up to 150 keV for this source, enabling effective constraint of the PL and reflection components, and the lower energy limit of 1 keV for Insight-HXMT ensures accurate modeling of the disk component. Moreover, unlike the Swift data used by \cite{2019Tao}, the spectra are not prone to distortion due to pile-up effects as Insight-HXMT does not suffer from this issue.

Insight-HXMT observed GRS 1716--249 twice during the 2016--2017 outburst. According to the spectral classification of \cite{2019Bassi}, both observations are in the hard intermediate state. NuSTAR observed the source three times in the hard intermediate state. The second observation by Insight-HXMT (obsID P0114335002) is strictly simultaneous with the third NuSTAR observation (obsID 90301007002), which started on July 28, 2017, and ended on July 30, 2017. The simultaneous observations by Insight-HXMT and NuSTAR are indicated on the outburst light curve obtained from MAXI/GSC (see Fig.~\ref{fig:lc}). The effective exposure times for Insight-HXMT/LE and NuSTAR/FPMA are 26\,ks and 89\,ks, respectively (see Table~\ref{tab:obs}). Since the observation span exceeds two days and the source's luminosity, that is, accretion rate, changes drastically during this period, we split the observations of Insight-HXMT and NuSTAR into two data sets (see Fig.~\ref{fig:lc2}) to more accurately measure the spin. The information on the divided data is provided in Table~\ref{tab:datasets}.

\subsection{Insight-HXMT}
\label{sec:hxmt}
We perform data reduction using the Insight-HXMT Data Analysis Software ({\tt HXMTDAS} {\tt v2.05}\footnote{\url{http://hxmten.ihep.ac.cn/software.jhtml}}) and the calibration database files ({\tt CALDB v2.06}). 
To determine the good time intervals, we established the following criteria: (a) pointing offset angle should be less than 0.04$^{\circ}$; (b) Earth elevation angle should be greater than 10$^{\circ}$; (c) there should be a minimum time interval of 300\,s from the crossing of the South Atlantic Anomaly region; (d) the geomagnetic cutoff rigidity should exceed 8\,GV. The background for the Low Energy (LE), Medium Energy (ME), and High Energy (HE) telescopes were generated using the scripts {\tt lebkgmap}, {\tt mebkgmap}, and {\tt hebkgmap}, respectively, based on the Insight-HXMT background models \citep{Liao2020a, Guo2020, Liao2020b}. The response files for LE, ME, and HE were generated using {\tt lerspgen}, {\tt merspgen}, and {\tt herspgen}, respectively. Following the recommendation of the Insight-HXMT calibration group, the combined spectra are rebinned as follows: The spectra of LE, ME and HE are respectively rebinned with 1000, 800 and 600 counts per bin at least. Additionally, systematic errors of 1\%, 1\%, and 2\% are applied to the LE, ME, and HE spectra, respectively.

\subsection{NuSTAR}
The {\tt nupipeline} routine of {\tt NuSTARDAS v1.9.7} in {\tt HEASoft v6.29} with {\tt CALDB v20211115}, is employed to process the cleaned event files. The {\tt nuproducts} tool is then used to extract the source events, by adopting a circular region surrounding the source with a radius of $80^{\prime \prime}$ to optimize the signal-to-noise ratio of the spectra \footnote{The script can be found in \url{https://github.com/NuSTAR/nustar-gen-utils/tree/3a603ca820a93c81414a298fd90d2e5a05f5e24a/notebooks}.}. The corresponding background extraction region is a nearby source-free circle with a radius of $80^{\prime \prime}$. The spectra are rebinned with 50 counts per bin at least.
%__________________________________________________________________

%by here
\section{Analysis and results}
\label{sec:res}

%%%%%%%%%%%%%%%%%%%%%%%%%%%%%%%TABLE%%%%%%%%%%%%%%%%%%%%%%%%%%%%%
\begin{figure}
    \centering
    \includegraphics[scale=0.5]{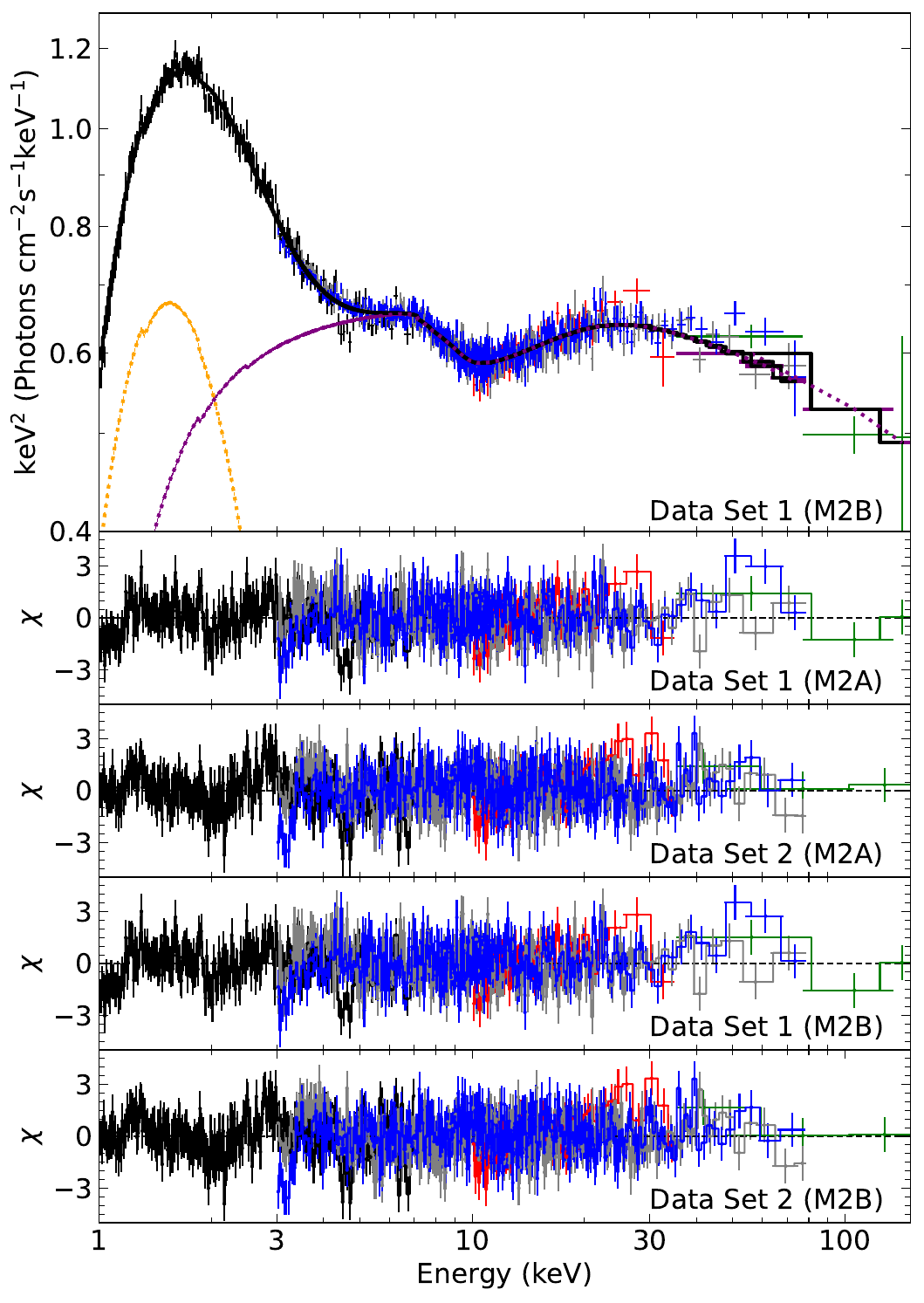}\\
    \caption{Spectra (black for Insight-HXMT/LE, red for Insight-HXMT/ME, green for Insight-HXMT/HE, gray for NuSTAR/FPMA and blue for NuSTAR/FPMB), model components of {\tt M2B}, and spectral residuals of {\tt M2A} and {\tt M2B}. The black solid line is the total model fitted to the data, and the yellow and purple dotted lines show the {\tt kerrbb2} and {\tt relxill} spectral components, respectively. The models are plotted based on the best-fit parameters obtained from Insight-HXMT.}
    \label{fig:fitting}
\end{figure}
%%%%%%%%%%%%%%%%%%%%%%%%%%%%%%%%TABLE%%%%%%%%%%%%%%%%%%%%%%%%%%%%%%%%%%%%%%%
\begin{table*}
 \renewcommand\arraystretch{1.5}
 
 \centering
\caption{Joint fitting parameters of Insight-HXMT and NuSTAR with {\tt M1} ({\tt constant*tbabs*(kerrbb+relxill))}. }
 \label{tab:table3}
 \begin{tabularx}{0.8\textwidth}{@{}>{\centering\arraybackslash}X|>{\centering\arraybackslash}X>{\centering\arraybackslash}X|>{\centering\arraybackslash}X>{\centering\arraybackslash}X@{}}
 \hline
 \hline
    &  \multicolumn{2}{c|}{{\tt M1A} ($D$ = 2.4\,kpc)}  &  \multicolumn{2}{c}{{\tt M1B} ($D$ = 6.9\,kpc)}  \\
\hline
 Parameter &    Data Set 1 &  Data Set 2   &    Data Set 1 &  Data Set 2     \\
 \hline
  & \multicolumn{4}{c}{\tt TBABS} \\
$N_{\rm H}^{\bigstar}$
  &  \multicolumn{2}{c|}{$0.758_{-0.016}^{+0.009}$} &  \multicolumn{2}{c}{$0.696_{-0.015}^{+0.006}$}  
\\
 \hline
   & \multicolumn{4}{c}{\tt KERRBB} \\
$\dot{M}$
& $14.4\pm0.5$
  &  $11.9\pm0.4$
  &  $262_{-15}^{+10}$
  &  $212_{-12}^{+8}$ 
  \\
$f_{\rm h}$
& $1.319_{-0.019}^{+0.069}$
  &  $1.29_{-0.02}^{+0.07}$
  &  $1.41_{-0.04}^{+0.08}$
  &  $1.40_{-0.04}^{+0.08}$
\\
\hline
    &\multicolumn{4}{c}{\tt RELXILL} \\
$q_{\rm in}$
&  $>9.35$
  &  $>9.73$
  &  $5.2_{-1.0}^{+0.8}$
  &  $>7.4$
\\
$R_{\rm b}$
&  $1.68^{+0.04}_{-0.11}$
  &  $1.75^{+0.06}_{-0.11}$
  &  $39.3^{+55.8}_{-29.4}$
  &  $4.6^{+10.8}_{-1.6}$
\\
$a_*^{\bigstar}$
  &  \multicolumn{2}{c|}{$0.989_{-0.005}^{+0.002}$}  &  \multicolumn{2}{c}{$0.44_{-0.08}^{+0.07}$}  
     \\
$i^{\bigstar}$ 
  &  \multicolumn{2}{c|}{$43.8_{-1.3}^{+0.6}$}  &  \multicolumn{2}{c}{$43.6_{-1.1}^{+0.5}$}  
    \\
$\rm \Gamma$
& $1.939\pm0.012$
  &  $1.892_{-0.013}^{+0.016}$
  &  $1.958_{-0.016}^{+0.009}$
  &  $1.900_{-0.018}^{+0.006}$ \\
log~$\xi$
& $3.93\pm0.07$
  &  $3.77_{-0.11}^{+0.07}$
  &  $3.86_{-0.08}^{+0.10}$
  &  $3.81_{-0.05}^{+0.10}$ \\
$A_{\rm Fe}^{\bigstar}$
  &  \multicolumn{2}{c|}{$>8.3$}  &  \multicolumn{2}{c}{$5.0_{-0.3}^{+0.4}$}    \\
$E_{\rm cut}$ 
& $>877.2$
  &  $840.2_{-118.3}^{+117.0}$
  &  $>950.4$
  &  $>938.8$ \\
$R_{\rm f}$
& $0.68_{-0.13}^{+0.06}$
  &  $0.58_{-0.09}^{+0.07}$
  &  $0.40_{-0.05}^{+0.08}$
  &  $0.36_{-0.03}^{+0.08}$ \\
norm 
& $7.2\pm0.4$  &  $7.7\pm0.4$  &  $7.8_{-0.6}^{+0.4}$  &  $7.9_{-0.6}^{+0.2}$
  \\
\hline
Flux  & 1.19 & 1.06 & 1.11  & 0.98
\\
\hline
$\chi ^2$/dof  & \multicolumn{2}{c|}{6342.6/6356 = 0.998} & \multicolumn{2}{c}{6404.8/6356 = 1.008}   \\
\hline
\end{tabularx}
\begin{tablenotes}
\item Notes. All errors are quoted at the 90\% confidence level. The probability distributions of parameters for {\tt M1B} obtained from Data Set 1 through MCMC are presented in Fig.~\ref{fig:corner1}; $^{\bigstar}$ indicates that the parameters between different data sets are linked; $N_{\rm H}$ is the X-ray absorption column density in units of $10^{22}$~atoms~$\rm cm^{-2}$; $\dot{M}$ is the effective mass accretion rate of the disk in units of $10^{15}~\rm g~\rm s^{-1}$; $f_{\rm h}$ is the spectral hardening factor; $q_{\rm in}$ is the inner emissivity index; $R_{\rm b}$ represents the transition radius between the inner and outer emissivity indices, measured in $R_{\rm ISCO}$ units, and is capped at 100\,$R_{\rm ISCO}$; $a_*$ is the BH spin; $i$ is the disk inclination angle in units of deg; $\Gamma$ is the photon index of the incident cutoff PL spectrum; $\xi$ is the ionization parameter of the accretion disk in units of $\rm erg~cm~s^{-1}$; $A_{\rm Fe}$ is the iron abundance of the accretion disk in units of solar abundance; $E_{\rm cut}$ is the cutoff energy in units of keV; $R_{\rm f}$ is the reflection fraction; $\rm norm$ is the normalization of {\tt relxill} in units of $10^{-3}$; Flux is the 0.1--100\,keV unabsorbed fluxes in units of $\rm 10^{-8}~erg~cm^{-2}~s^{-1}$.
\end{tablenotes}
\end{table*}
%%%%%%%%%%%%%%%%%%%%%%%%%%%%%%%%%%%%%%%%%
%KERRBB2------------------------------------------------------------------------------
\begin{table*}
 \renewcommand\arraystretch{1.5}
 
 \centering
\caption{Joint fitting parameters of Insight-HXMT and NuSTAR with {\tt M2} ({\tt constant*tbabs*(kerrbb2+relxill))}. }
 \label{tab:table4}
 \begin{tabularx}{0.8\textwidth}{@{}>{\centering\arraybackslash}X|>{\centering\arraybackslash}X>{\centering\arraybackslash}X|>{\centering\arraybackslash}X>{\centering\arraybackslash}X@{}}
 \hline
 \hline
    &  \multicolumn{2}{c|}{{\tt M2A} ($D$ = 2.4\,kpc)}  &  \multicolumn{2}{c}{{\tt M2B} ($D$ = 6.9\,kpc)}  \\
\hline
 Parameter &    Data Set 1 &  Data Set 2   &    Data Set 1 &  Data Set 2     \\
 \hline
  & \multicolumn{4}{c}{\tt TBABS} \\
$N_{\rm H}^{\bigstar}$
  &  \multicolumn{2}{c|}{$0.721_{-0.014}^{+0.008}$} &  \multicolumn{2}{c}{$0.695_{-0.011}^{+0.009}$}  
\\
 \hline
   & \multicolumn{4}{c}{\tt KERRBB2} \\
$\dot{M}$
& $15.4_{-0.5}^{+0.4}$
  &  $12.2_{-0.4}^{+0.3}$
  &  $258_{-8}^{+5}$
  &  $208_{-7}^{+4}$
  \\
\hline
    &\multicolumn{4}{c}{\tt RELXILL} \\
$q_{\rm in}$
&  $>8.2$
  &  $>9.3$
  &  $5.2_{-0.8}^{+0.7}$
  &  $9.5_{-2.2}^{+0.4}$
\\
$R_{\rm b}$
&  $1.46^{+0.04}_{-0.03}$
  &  $1.53\pm0.04$
  &  $32.3^{+61.6}_{-24.7}$
  &  $7.2^{+50.4}_{-4.0}$
\\
$a_*^{\bigstar}$
  &  \multicolumn{2}{c|}{$0.972_{-0.005}^{+0.004}$}  &  \multicolumn{2}{c}{$0.464_{-0.007}^{+0.016}$}  
     \\
$i^{\bigstar}$ 
  &  \multicolumn{2}{c|}{$38.8_{-0.6}^{+1.6}$}  &  \multicolumn{2}{c}{$43.8_{-0.8}^{+0.5}$}  
    \\
$\rm \Gamma$
& $1.89_{-0.02}^{+0.03}$
  &  $1.875_{-0.006}^{+0.008}$
  &  $1.951_{-0.008}^{+0.006}$
  &  $1.899_{-0.006}^{+0.005}$ \\
log~$\xi$
& $4.14_{-0.06}^{+0.04}$
  &  $3.91_{-0.06}^{+0.03}$
  &  $3.90_{-0.06}^{+0.05}$
  &  $3.81_{-0.05}^{+0.03}$ \\
$A_{\rm Fe}^{\bigstar}$
  &  \multicolumn{2}{c|}{$>9.3$}  &  \multicolumn{2}{c}{$5.0_{-0.3}^{+0.4}$}    \\
$E_{\rm cut}$ 
& $>826.2$
  &  $673.9_{-102.2}^{+86.9}$
  &  $>854.0$
  &  $>904.1$ \\
$R_{\rm f}$
& $0.65_{-0.15}^{+0.12}$
  &  $0.41_{-0.04}^{+0.03}$
  &  $0.44\pm0.04$
  &  $0.36\pm0.02$ \\
norm 
& $6.3_{-0.6}^{+0.7}$  &  $7.5_{-0.2}^{+0.3}$  &  $7.5_{-0.3}^{+0.2}$  &  $7.9\pm0.2$
  \\
\hline
Flux  & 1.12 & 1.00 & 1.11  & 0.98
\\
\hline
$\chi ^2$/dof  & \multicolumn{2}{c|}{6372.2/6358 = 1.002} & \multicolumn{2}{c}{6406.9/6358 = 1.008}   \\
\hline
\end{tabularx}
\begin{tablenotes}
\item Notes. All errors are quoted at the 90\% confidence level. The probability distributions of parameters for {\tt M2A} and {\tt M2B} obtained from Data Set 1 through MCMC are presented in Figs.~\ref{fig:corner2} and \ref{fig:corner3}, respectively.
\end{tablenotes}
\end{table*}
%%%%%%%%%%%%%%%%%%%%%%%%%%%%%%%%TABLE%%%%%%%%%%%%%%%%%%%%%%%%%%%%%%%%%%%%%%%
%\begin{landscape}
\begin{table*}
 \renewcommand\arraystretch{1.5}
 \small
 \centering
\caption{Fitting results of partial parameters with {\tt M2} by assuming different distances. }
 \label{tab:table5}
 
 \begin{tabularx}{0.95\textwidth}{@{}>{\centering\arraybackslash}X>{\centering\arraybackslash}X>{\centering\arraybackslash}X>{\centering\arraybackslash}X>{\centering\arraybackslash}X>{\centering\arraybackslash}X>{\centering\arraybackslash}X>{\centering\arraybackslash}X@{}}
 \hline
 \hline
 $D$ (kpc) &    2.4 ({\tt M2A})  &  3  &  4  &  5  &  6  &  6.9 ({\tt M2B})  &  8      \\
$M=6.4M_{\rm \odot}$ &  &  &  &  &  &  &  \\
 \hline
 $a_{*}$  & $0.972_{-0.005}^{+0.004}$ ($0.946_{-0.008}^{+0.012}$) & $0.958_{-0.011}^{+0.007}$ ($0.847_{-0.005}^{+0.013}$) & $0.721_{-0.004}^{+0.019}$ & $0.634\pm0.014$ & $0.530_{-0.007}^{+0.018}$ & $0.464_{-0.007}^{+0.016}$ & $0.393_{-0.015}^{+0.016}$ \\
 $i$ ($^{\circ}$) & $38.8_{-0.6}^{+1.6}$ ($44.1_{-1.5}^{+1.3}$) & $35.1_{-0.7}^{+2.0}$ ($47.0_{-1.1}^{+0.7}$) & $46.2_{-1.1}^{+0.7}$ & $44.7_{-0.8}^{+0.9}$ & $44.4_{-0.7}^{+0.5}$ & $43.8_{-0.8}^{+0.5}$ & $43.0_{-0.8}^{+0.5}$ 
 \\
 $\dot{M}$  &  $15.4_{-0.5}^{+0.4}$ ($16.4_{-0.9}^{+0.6}$)   &  $24.7\pm1.0$ ($33.2_{-1.4}^{+0.7}$)  &  $70.8_{-3.3}^{+0.8}$   &  $119_{-3}^{+4}$   &  $188_{-7}^{+3}$   &  $258_{-8}^{+5}$   &  $354\pm9$ 
 \\
 $A_{\rm Fe}$  &  $>9.3$   (5)  &  $>9.3$   (5)  &  $5.0_{-0.2}^{+0.6}$   &  $5.0_{-0.2}^{+0.7}$   &  $5.0_{-0.3}^{+0.4}$   &  $5.0_{-0.3}^{+0.4}$   &  $5.0_{-0.3}^{+0.4}$ 
 \\  \hline
 $\chi^2$ [dof=6358 (6359)] & 6372.2 (6430.3) & 6410.9 (6426.4) & 6412.6  & 6409.0 & 6406.8 & 6406.9 &  6411.4
\\
\hline
\end{tabularx}
\begin{tablenotes}
\item Notes. All errors are quoted at the 90\% confidence level. The joint fitting results of BH spin, disk inclination, accretion rate of Data Set 1, iron abundance, and goodness of fitting at different distances with {\tt M2} are listed in the table (values without parentheses). For distances of 2.4 and 3 kpc, besides the results with the iron abundance freely fitted, we also include the fitting parameters set to a fixed iron abundance of 5 (values in parentheses).
\end{tablenotes}
\end{table*}

%%%%%%%%%%%%%%%%%%%%%%%%%%%%%%%%%%%%%%
To obtain more reliable results, we performed a joint fit of the spectra from Data Sets 1 and 2. The energy bands used for Insight-HXMT data are 1--7\,keV for LE, 10--35\,keV for ME, and 35--150\,keV for HE. For NuSTAR, we select the energy band of 3--79\,keV. The spectral fitting is conducted using {\tt XSPEC v12.12.0} \citep{1996Arnaud}. The abundances are set to WILM \citep{2000Wilms}, and the cross-sections are set to VERN \citep{1996Verner}. All errors are estimated via the Markov chain Monte Carlo algorithm (MCMC) with a length of 600000.

Due to the source being in the hard intermediate state \citep{2019Bassi}, where the spectra exhibit prominent disk emission and reflection features \citep{2019Tao}, we perform a joint fitting of the spectra using both a continuum and reflection spectral model. The model selected for the fitting is {\tt tbabs*(kerrbb+relxill)}. Additionally, a multiplicative constant model ({\tt constant}) is included to account for the normalization discrepancies between different telescopes. Therefore, our fitting model is {\tt M1} = {\tt constant*tbabs*(kerrbb+relxill)}, as shown in Table~\ref{tab:models}. The {\tt kerrbb} model is a multi-temperature blackbody model that describes the spectrum emitted by a geometrically thin, steady-state accretion disk around a Kerr black hole, taking into account full relativistic effects \citep{2005Li}. The {\tt relxill}\footnote{Version 2.0 of the relxill model is used in this work.} model is a relativistic disk coronal reflection model that describes the reflection produced by the corona (with a cutoff PL spectrum) illuminating the inner regions of the accretion disk \citep{2014Dauser,2014Garcia}. In the {\tt kerrbb} model, we fix the BH mass at 6.4\,$M_{\odot}$ \citep{2023Casares} and the normalization at 1, and link the BH spin ($a_*$) and the disk inclination ($i$) to that of {\tt relxill}. In the {\tt relxill} model, we fix the inner radius of the accretion disk ($R_{\rm in}$) to $-1$ since the disk inner edge is at the ISCO \citep{2019Bassi}, and presume that within the range from $R_{\rm in}$ to a specific break radius ($R_{\rm b}$), the emissivity is defined by a single inner emissivity index ($q_{\rm in}$), and beyond $R_{\rm b}$, it generally adopts the standard $r^{-3}$ ($q_{\rm out}=3$) profile. For different data sets, some parameters are linked and allowed to vary freely, including the X-ray absorption column density ($N_{\rm H}$), $a_*$, $i$, and iron abundance ($A_{\rm Fe}$). Other parameters are independent between different data sets, such as the mass accretion rate ($\dot{M}$) and the spectral hardening factor $f_{\rm h}$ in {\tt kerrbb}, and $q_{\rm in}$, $R_{\rm b}$, the photon index ($\Gamma$), ionization parameter (log~$\xi$), cutoff energy of the incident spectrum ($E_{\rm cut}$), reflection factor ($R_{\rm f}$), and normalization in {\tt relxill}. To determine the BH spin under the most recent measured distance and investigate the influence of distance on spin measurement, we perform two sets of fits: fixing the distance $D$ in the {\tt kerrbb} at 2.4\,kpc ({\tt M1A}; See Table~\ref{tab:models}) and 6.9\,kpc ({\tt M1B}), respectively.

The two fittings at different distances yield similar goodness of fit, and their detailed fitting parameters are listed in Table~\ref{tab:table3}. In both models {\tt M1A} and {\tt M1B}, the accretion rate shows a gradual decrease from Data Set 1 to Data Set 2, which aligns with the light curve's variability; both models yield $N_{\rm H}\sim$ 0.7, $i\sim 44^{\circ}$, $\Gamma\sim 1.9$, log~$\xi \sim 3.8$, and $E_{\rm cut}\gtrsim$ 900\,keV, indicating that these parameters do not exhibit strong distance dependence. The spin fitting results exhibit noteworthy changes, where an increase in distance transitions high spin values to moderate ones. The relationship between $a_*$ and $f_{\rm h}$ shown in Fig.~\ref{fig:corner1} demonstrates an inverse correlation, in agreement with the findings of \cite{2021Salvesen}. This suggests that $f_{\rm h}$ has a significant impact on spin fitting. To reduce this impact and achieve reliable spin measurements, we have substituted {\tt kerrbb} in {\tt M1} with {\tt kerrbb2} \citep{2006McClintock_kerrbb2}. Unlike {\tt Kerrbb}, {\tt Kerrbb2} incorporates the spectral hardening effect using two search tables for $f_{\rm h}$, each corresponding to different viscosity parameters: $\alpha$ = 0.01 and 0.1. These look-up tables are generated via {\tt bhspec} \citep{2005Davis}, relying on non-LTE atmosphere models within an $\alpha$-viscosity framework. Other features of {\tt Kerrbb2}, such as Doppler boosting, gravitational redshift, and returning radiation, remain consistent with {\tt kerrbb}. The updated model is now {\tt M2} = {\tt constant*tbabs*(kerrbb2+relxill)}, as shown in Table~\ref{tab:models}. We chose $\alpha$ = 0.1 \citep{2011Steiner} in {\tt M2} and kept the other parameter settings the same as in {\tt M1}, such as keeping the black hole mass fixed at 6.4\,$M_{\rm \odot}$. Additionally, similar to {\tt M1}, we fixed the distance parameter at two specific values: $D$ = 2.4\,kpc ({\tt M2A}; See Table~\ref{tab:models}) and $D$ = 6.9\,kpc ({\tt M2B}), respectively.

%The primary difference between the two models is that {\tt kerrbb} fits the hardening factor, while {\tt kerrbb2} requires the selection of the viscosity parameter. 

%\textbf{Compared to {\tt M1}, the difference in {\tt M2} is that {\tt Kerrbb2} incorporates the effect of spectral hardening through a pair of look-up tables for $f_{\rm h}$ corresponding to two viscosity parameter values: $\alpha$ = 0.01 and 0.1. These tables are computed using {\tt bhspec} \citep{2005Davis} based on non-LTE atmosphere models within an $\alpha$-viscosity framework. Other features of {\tt Kerrbb2}, such as Doppler boosting, gravitational redshift, and returning radiation, remain consistent with {\tt kerrbb}. The main distinction between the two models lies in the parameterization: {\tt M1} involves fitting the hardening factor in {\tt kerrbb}, while {\tt M2} requires selecting the viscosity parameter in {\tt kerrbb2}.

The fitting results with {\tt M2} are showed in Table~\ref{tab:table4} and Fig.~\ref{fig:fitting}. Compared to {\tt M1}, the fitting results of {\tt M2} did not exhibit notable differences, with {\tt M2A} leading to a high spin value of $a_{*} \sim$ 0.97, and {\tt M2B} resulting in a lower spin value of $a_{*} \sim$ 0.46. These discrepancies in spin are attributed to the different distance assumptions. To investigate the influence of distance variations on spin measurements, by maintaining a constant BH mass, we conduct additional spectral fits using {\tt M2} with different distance parameters. The resulting spin values, inclination angles, accretion rate of Data Set 1, iron abundance, and goodness of fit are detailed in Table~\ref{tab:table5}. The analysis revealed that the best-fit iron abundance exhibits two distinct sets of values: for distances of 2.4 and 3\,kpc, we observe very high values of $A_{\rm Fe}>9.3$; in contrast, for distances between 4 and 8\,kpc, the iron abundance stabilizes around $A_{\rm Fe} \sim$ 5. The latter $A_{\rm Fe}$ is similar to the values measured for other BH binaries, such as GX 339--4 with $A_{\rm Fe} = 5\pm 1$ \citep{2015Garcia}, V404 Cyg with $A_{\rm Fe}\sim 5$ \citep{2017Walton}, and Cyg X--1 with $A_{\rm Fe} = 4.7\pm 0.1$ \citep{2015Parker}. To examine the effect of different $A_{\rm Fe}$, we fixed $A_{\rm Fe}$ at 5 for distances of 2.4 and 3\,kpc during the fitting process. The corresponding results are shown in Table~\ref{tab:table5} in parentheses, where $a_*$ shows a slight decrease and $i$ becomes more consistent with the cases of larger distances. Our analysis demonstrates a significant decrease in the spin with increasing distance. This emphasizes the importance of accurate distance determination when using continuum spectrum fitting for BH spin measurements. Any inaccuracies in the distance estimation can lead to deviations in the measured spin values, even if the reflection component modeling is also employed.

\section{Discussion}
\label{sec:dis}
%%%%%%%%%%%%%%%%%%%%%%%%%%%%%%%%%%%%%%%%%%%%%%%%%%%%%%%%%%%%%%%%
\begin{figure}
    \centering
    \includegraphics[scale=0.48]{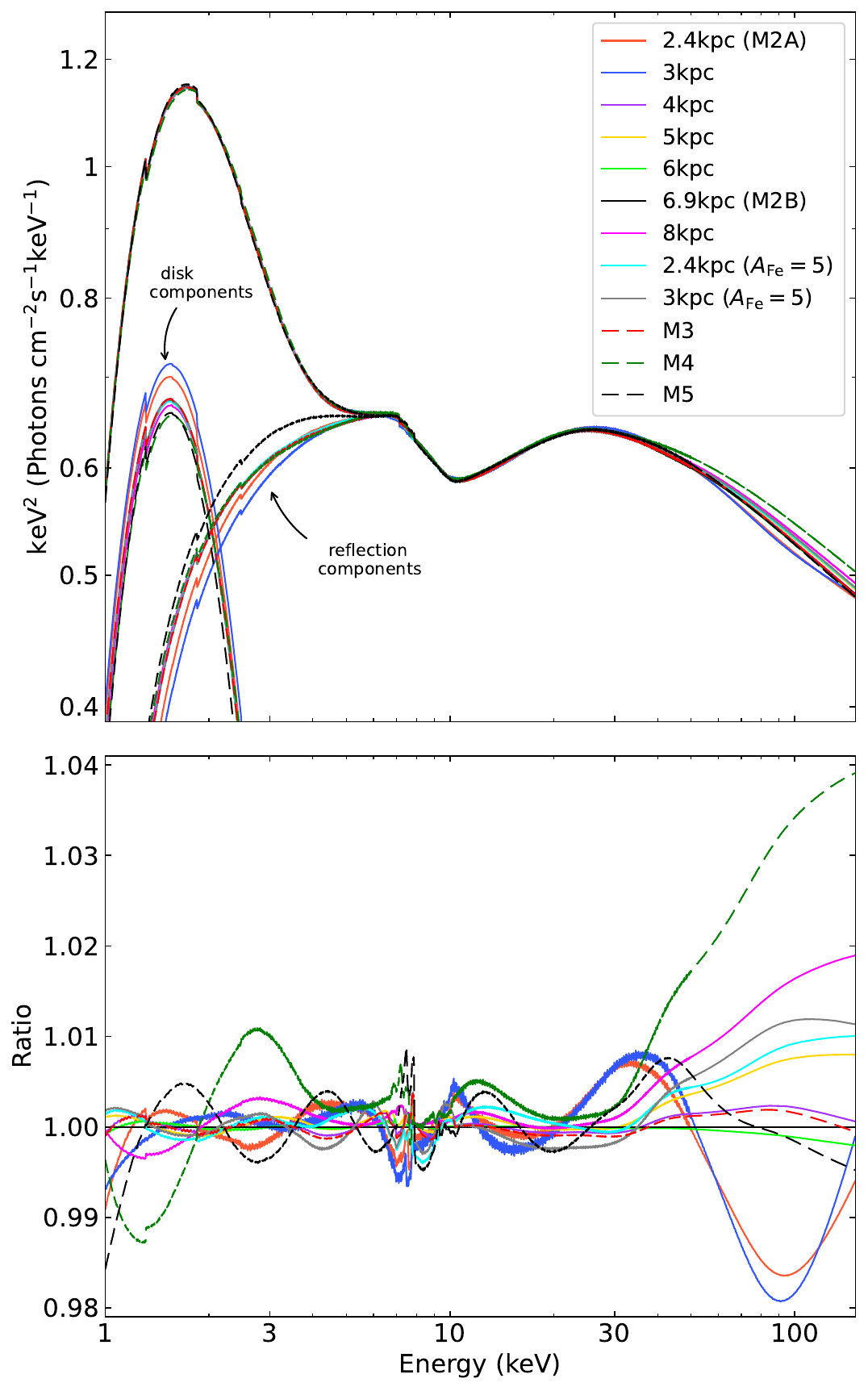}\\
    \caption{Comparison diagram of different models. Top panel: Different models and their disk components and reflection components. Bottom panel: The ratio of each model relative to {\tt M2B}.}
    \label{fig:model_ratio}
\end{figure}
%%%%%%%%%%%%%%%%%%%%%%%%%%%%%%%%TABLE%%%%%%%%%%%%%%%%%%%%%%%%%%%%%%%%%%%%%%%
%%________________________________________________________________________
%DISKBB
\begin{table}
 \renewcommand\arraystretch{1.5} 
 \centering
\caption{Joint fitting parameters of Insight-HXMT and NuSTAR with {\tt Tbabs*(diskbb+relxill)}.}
 \label{tab:table6}
 \begin{minipage}{.48\textwidth}
 \begin{tabularx}{\textwidth}{@{}>{\centering\arraybackslash}X>{\centering\arraybackslash}X>{\centering\arraybackslash}X@{}}
 \hline
 \hline
 Parameter &    Data Set 1 &  Data Set 2     \\
 \hline
  & \multicolumn{2}{c}{\tt TBABS} \\
$N_{\rm H}^{\bigstar}$
  &  \multicolumn{2}{c}{$0.699_{-0.019}^{+0.008}$} 
\\
 \hline
   & \multicolumn{2}{c}{\tt DISKBB} \\
$T_{\rm in}$
& $0.425_{-0.003}^{+0.008}$
  &  $0.3974_{-0.0014}^{+0.0080}$
  \\
$\rm Norm$
& $5454_{-526}^{+215}$
  &  $5752_{-631}^{+171}$ \\
\hline
    &\multicolumn{2}{c}{\tt RELXILL} \\
$q_{\rm in}$
&  $9.0_{-1.0}^{+0.7}$
  &  $>9.5$  \\
$R_{\rm b}$
&  $1.9\pm0.2$
  &  $1.9\pm0.2$  \\
$a_*^{\bigstar}$
  &  \multicolumn{2}{c}{$0.9943_{-0.0044}^{+0.0005}$}  
     \\
$i^{\bigstar}$ 
  &  \multicolumn{2}{c}{$47.6_{-1.5}^{+1.8}$}  
    \\
$\rm \Gamma$
& $1.969_{-0.019}^{+0.009}$
  &  $1.920\pm0.010$ \\
log~$\xi$
& $3.69_{-0.08}^{+0.13}$
  &  $3.56_{-0.07}^{+0.05}$ \\
$A_{\rm Fe}^{\bigstar}$
  &  \multicolumn{2}{c}{$8.7_{-2.5}^{+0.9}$}  
  \\
$E_{\rm cut}$ 
& $>786.7$
  &  $>834.4$ \\
$R_{\rm f}$
& $0.69_{-0.10}^{+0.11}$
  &  $0.75_{-0.14}^{+0.02}$ \\
norm 
& $7.9_{-0.6}^{+0.2}$  &  $7.80_{-0.18}^{+0.26}$
  \\
\hline
Flux  & 1.11 & 0.98
\\
\hline
$\chi ^2$/dof  & \multicolumn{2}{c}{6352.3/6356 = 0.999}  \\
\hline

\end{tabularx}
\begin{tablenotes}
\item Notes. All errors are quoted at the 90\% confidence level. The definition of each parameter is the same as that in Table~\ref{tab:table3}. 
\end{tablenotes}
\end{minipage}

\end{table}

%%%%%%%%%%%%%%%%%%%%%%%%%%%%%%%%%%%%%%%%%%%

In this paper, we perform a joint fitting of the simultaneous spectra of GRS 1716--249 observed by Insight-HXMT and NuSTAR. The data used are obtained during the softest phase of the outburst, where the disk component is most prominent and the inner edge of the disk is located at the ISCO \citep{2019Bassi}. Additionally, the presence of significant reflection features in the spectra allows us to measure the spin and inclination of the black hole using a combined fitting method of the continuum and reflection components.

\subsection{System Parameters with Updated Distance}

The fitting of the continuum is intrinsically linked to the distance, as altering the distance will impact the determination of the inner disk radius, consequently influencing the constraint on the spin. By fitting the spectra using {\tt M2} with varying distance parameters, we obtain distinct spin results (refer to Table~\ref{tab:table5}). Notably, assuming a distance of 2.4\,kpc, we obtain a near-extreme spin, aligning with the findings in \cite{2019Tao}. However, assuming a distance of 6.9\,kpc based on the recent works of \cite{2022Saikia} and \cite{2023Casares}, resulted in a moderate spin of $a_{*} \sim 0.46$, implying that this source is not a rapidly rotating BH. This suggests that the spin was previously overestimated under the prior distance assumption. The disk inclination, even with varying distance assumptions, remains nearly constant around $43-47^{\circ}$, which is within the error margin of the orbital inclination of $61\pm15^{\circ}$ (with 68\% confidence) as reported by \cite{2023Casares}.

In the joint fitting of the continuum and reflection spectra, the distance measurement primarily affects the fitting of the continuum spectrum model ({\tt kerrbb2}). Therefore, to investigate the impact of distance changes on the model, we plotted the best-fit {\tt M2} with its disk component {\tt Kerrbb2} and reflection component {\tt relxill} at different distances (shown as solid lines in Fig.~\ref{fig:model_ratio}). The variation in distance did not result in significant changes in the total models, with differences of less than 2\% in each energy band (see Fig.~\ref{fig:model_ratio} bottom panel). Importantly, we found that the energy corresponding to the peak flux of the disk component ($E_{\rm peak}^{\rm disk}$) did not show significant changes. Since the observed flux is fixed, an increase in distance leads to an increase in the accretion rate, as indicated by the parameter $\dot{M}$ in Table~\ref{tab:table5}. According to accretion disk theory \citep{1973Shakura}, when other parameters remain constant, an increase in the accretion rate causes $E_{\rm peak}^{\rm disk}$ to shift towards higher energies, while a decrease in spin leads to an increase in $R_{\rm in}$ (=$R_{\rm ISCO}$), resulting in a decrease in $E_{\rm peak}^{\rm disk}$. Therefore, to maintain a constant $E_{\rm peak}^{\rm disk}$, the model compensates for the increase in accretion rate due to larger distances by reducing the spin. This corresponds to the inverse correlation between $a_{*}$ and $\dot{M}$ shown in Figs.~\ref{fig:corner1}--\ref{fig:corner3}. 

Furthermore, when other parameters remain constant, an increase in BH mass will cause $E_{\rm peak}^{\rm disk}$ to shift towards lower energies. In this case, the model compensates for the mass increase by increasing the spin. Therefore, within the constraints of distance and mass provided by \cite{2023Casares}, setting the lower limit (upper limit) of distance and the upper limit (lower limit) of mass yields the upper (lower) limit of the spin. When fixing the distance at 5.8\,kpc and the BH mass at $9.6~M_{\rm \odot}$ ({\tt M3}; See Table~\ref{tab:models}), we obtain a moderately high spin value of $a_{*}=0.757_{-0.011}^{+0.023}$, with a goodness of fit of $\chi ^2$/dof = 6416.4/6358 (see Fig.~\ref{fig:corner4} for the parameter probability distribution of Data Set 1). When fixing the distance at 8\,kpc and the BH mass at $4.4~M_{\rm \odot}$ ({\tt M4}; See Table~\ref{tab:models}), we obtain a near non-rotating black hole with $a_{*}=-0.008_{-0.013}^{+0.011}$, with a goodness of fit of $\chi ^2$/dof = 6499.9/6358. The red and green dashed lines in Fig.~\ref{fig:model_ratio} represent {\tt M3} and {\tt M4}. Therefore, considering the upper and lower limits of distance and mass provided by \cite{2023Casares}, there will be a wider range of spin with $a_{*}$ < 0.78.

The above discussion indicates that the variation in the measured spin values with changes in distance and mass is a result of the degeneracy in the model. A similar phenomenon was observed when measuring the spins of LMC X--1 and Cyg X--1 using continuum spectra \citep{2024Zdziarski}. Their analysis shows that the disk components dominate the spectra, so $E_{\rm peak}^{\rm disk}$ derived from fitting matches the energy of the peak flux seen in the spectra ($E_{\rm peak}^{\rm obs}$). Increasing the hardening factor or convolving with Comptonization components outside the disk components will shift the $E_{\rm peak}^{\rm disk}$ towards higher energies. To maintain $E_{\rm peak}^{\rm disk}$ corresponding as closely as possible to $E_{\rm peak}^{\rm obs}$, the remaining free parameters in the continuum spectrum, spin, and accretion rate, will decrease accordingly to counteract the effects of the hardening factor or Comptonization component.

Furthermore, the inconsistency in iron abundance at different distances as shown in Table~\ref{tab:table5} ($A_{\rm Fe}$ > 9.3 for $D$ = 2.4 and 3\,kpc; $A_{\rm Fe} \sim$ 5 for $D$ = 4--8\,kpc) primarily reflects not only the degeneracies among the parameters of the continuum model but also a certain level of competition between the continuum and reflection components, particularly in the low energy band of $\lesssim$ 2\,keV \citep{2018Garcia_a,2018Tomsick}. For instance, {\tt M2} with $D$ = 2.4 and 3\,kpc, as depicted in Fig.~\ref{fig:model_ratio}, exhibits a relatively higher proportion of disk component and a diminished reflection component compared to other models. The shape of the reflection spectrum in the low energy is significantly influenced by the configuration of the incident spectrum and the parameters of the disk as the reflector, leading to intertwined dependencies among various parameters in the overall model and rendering the parameter space exceptionally complicated. \cite{2024Zdziarski} reported local minimum iron abundance of $A_{\rm Fe} \lesssim$ 1 and global minimum values of $A_{\rm Fe} \sim$ 8 at $\Delta \chi^{2} \approx -0.6$ while fitting the spectra of LMC X-1 using continuum and reflection components ({\tt kerrbb2+reflkerr}), thereby corroborating this point. Therefore, for obtaining more rational and precise parameters during spectral fitting, a comprehensive analysis of the entire parameter space is imperative to assess potential biases introduced.

\subsection{Further Insights into Spin Measurements}

Given this, we attempted to replace {\tt kerrbb2} with {\tt diskbb}, which only has two parameters, disk temperature $T_{\rm in}$ in units of keV and Norm, without considering the impact of distance, and re-conduct the spectral fitting ({\tt M5} = {\tt constant*tbabs*(diskbb+relxill)}; See Table~\ref{tab:models}). The fitting results are shown in Table~\ref{tab:table6}, which are very close to the fitting results of {\tt M1A}. We obtain an extreme spin of $a_{*}\gtrsim 0.99$, which is similar to the results obtained by \cite{2023Draghis} using the reflection component fitting method, that is, using the model {\tt diskbb} combined with a reflection model. Furthermore, utilizing {diskbb} Norm $\approx$ 5000--6000 and $i \approx 48^{\circ}$ of {\tt M5}, and assuming a distance of $D$ = 5.8--8\,kpc, a hardening factor of $f_{\rm h}=1.3$ (from {\tt M1A}), and a boundary condition correction factor $\zeta = 0.412$, we obtain the inner disk radius $R_{\rm in}\approx$ 35--53\,km using the formula $R_{\rm in}=\sqrt{\rm Norm/\cos{i}}\times \zeta \times f_{\rm h}^{2} \times D/10{\rm kpc}$ \citep{1998Kubota}. Assuming the source mass of 4.4--9.6~$M_{\rm \odot}$ \citep{2023Casares}, the gravitational radius $R_{\rm g}$ is about 6.5--14.3\,km, thus {\tt M5} yields $R_{\rm in}\approx 2.4-8.2~R_{\rm g}$. By setting $R_{\rm in}=R_{\rm ISCO}$, the spin derived from the {\tt diskbb} component is $a_{*}\sim -0.72-0.88$, which is inconsistent with the result of extreme spin obtained from the reflection model.

Similar to GRS 1716--249, when constraining the spin of some other sources, such as LMC X-3 \citep{2023Draghis,2014Steiner}, H 1743--322 \citep{2023Draghis_apj,2012Steiner}, and MAXI J1820+070 \citep{2023Draghis_apj,2021Guan,2021Zhao}, results obtained solely from the reflection component fitting method yield close to extreme spin values, while the continuum spectrum fitting method yields low spins of $a_{*}\lesssim0.3$. The examples above demonstrate that when fitting the spectrum, using either the continuum or the reflection component fitting method separately resulted in different inner radii of the accretion disk, leading to different spin values and hence physically inconsistent results. The joint fitting method of the continuum and reflection components may help to avoid this issue by considering the jointly optimal parameter space during the fitting process.

%%-------------------------------------------------------------------------------
\section{Conclusion}
\label{sec:con}

In conclusion, we re-evaluated several key parameters of the black hole GRS 1716--249 utilizing simultaneous data from Insight-HXMT and NuSTAR. This paper, through the application of a combined fitting method for the continuum and reflection components and the integration of updated distance and black hole mass, found that the black hole has a moderate spin and a moderately inclined accretion disk. Given the source distance of 6.9\,kpc and the black hole mass of 6.4$M_{\rm \odot}$, $a_{*}$ is $0.464_{-0.007}^{+0.016}$ and $i$ is $43.8_{-0.8}^{+0.5}$$^\circ$ with 90\% confidence level. Taking into account the uncertainties in distance and black hole mass, the spin range extends with $a_{*}$ < 0.78.

%Our findings indicate that the measurement of black hole spin is influenced by the choice of model and the setting of model parameters, emphasizing the need for improvement in the spectral models. 

%__________________________________________________________________
\begin{acknowledgements}
We thank the anonymous referee for useful comments that have allowed us to improved this manuscript. This work made use of data from the Insight-HXMT mission, a project funded by China National Space Administration (CNSA) and the Chinese Academy of Sciences (CAS). This work is supported by the National Key R\&D Program of China (2021YFA0718500). We acknowledge funding support from the National Natural Science Foundation of China (NSFC) under grant Nos. 12122306, 12333007 and 12027803, the CAS Pioneer Hundred Talent Program Y8291130K2, the Strategic Priority Research Program of the Chinese Academy of Sciences under grant No. XDB0550300, the Scientific and technological innovation project of IHEP Y7515570U1 and the International Partnership Program of Chinese Academy of Sciences (Grant No.113111KYSB20190020) .
\end{acknowledgements}

%-------------------------------------------------------------------

\bibliographystyle{aa}
\bibliography{ref}
%-------------------------------------------------------------------
\begin{appendix}
\section{Definitions of various models}
\label{sec:app_models}

\begin{table*}
\centering
\renewcommand{\arraystretch}{1.5}
\caption{Various models and the setting of the black hole mass ($M$) and distance ($D$) parameters in the disk components.}
\label{tab:models}
\begin{tabularx}{0.65\textwidth}{@{}>{\arraybackslash}X@{}}
\hline
\hline
{\tt M1} = {\tt constant*tbabs*(kerrbb+relxill)} with $M$ = $6.4~M_{\rm \odot}$ \\
{\tt M1A} = {\tt M1} with $D$ = 2.4\,kpc \\
{\tt M1B} = {\tt M1} with $D$ = 6.9\,kpc \\
\hline
{\tt M2} = {\tt constant*tbabs*(kerrbb2+relxill)} with $M$ = $6.4~M_{\rm \odot}$ \\
{\tt M2A} = {\tt M2} with $D$ = 2.4\,kpc \\
{\tt M2B} = {\tt M2} with $D$ = 6.9\,kpc \\
\hline
{\tt M3} = {\tt constant*tbabs*(kerrbb2+relxill)} with $M$ = $9.6~M_{\rm \odot}$ and $D$ = 5.8\,kpc \\
\hline
{\tt M4} = {\tt constant*tbabs*(kerrbb2+relxill)} with $M$ = $4.4~M_{\rm \odot}$ and $D$ = 8\,kpc \\
\hline
{\tt M5} = {\tt constant*tbabs*(diskbb+relxill)} \\
\hline
\end{tabularx}
%\begin{tablenotes} 
%\item Notes. 
%\end{tablenotes} 
\end{table*}

\section{Probability distributions of the parameters from Data Set 1}
\label{sec:app}

%\begin{figure*}
%    \centering
%    \includegraphics[width=1\textwidth]{kerrbb_2.4kpc.pdf}
%    \caption{Probability distributions of the parameters for {\tt M1A} obtained from Data Set 1 through MCMC.}
%    \label{fig:corner0}
%\end{figure*}  

\begin{figure*}
    \centering
    \includegraphics[width=1\textwidth]{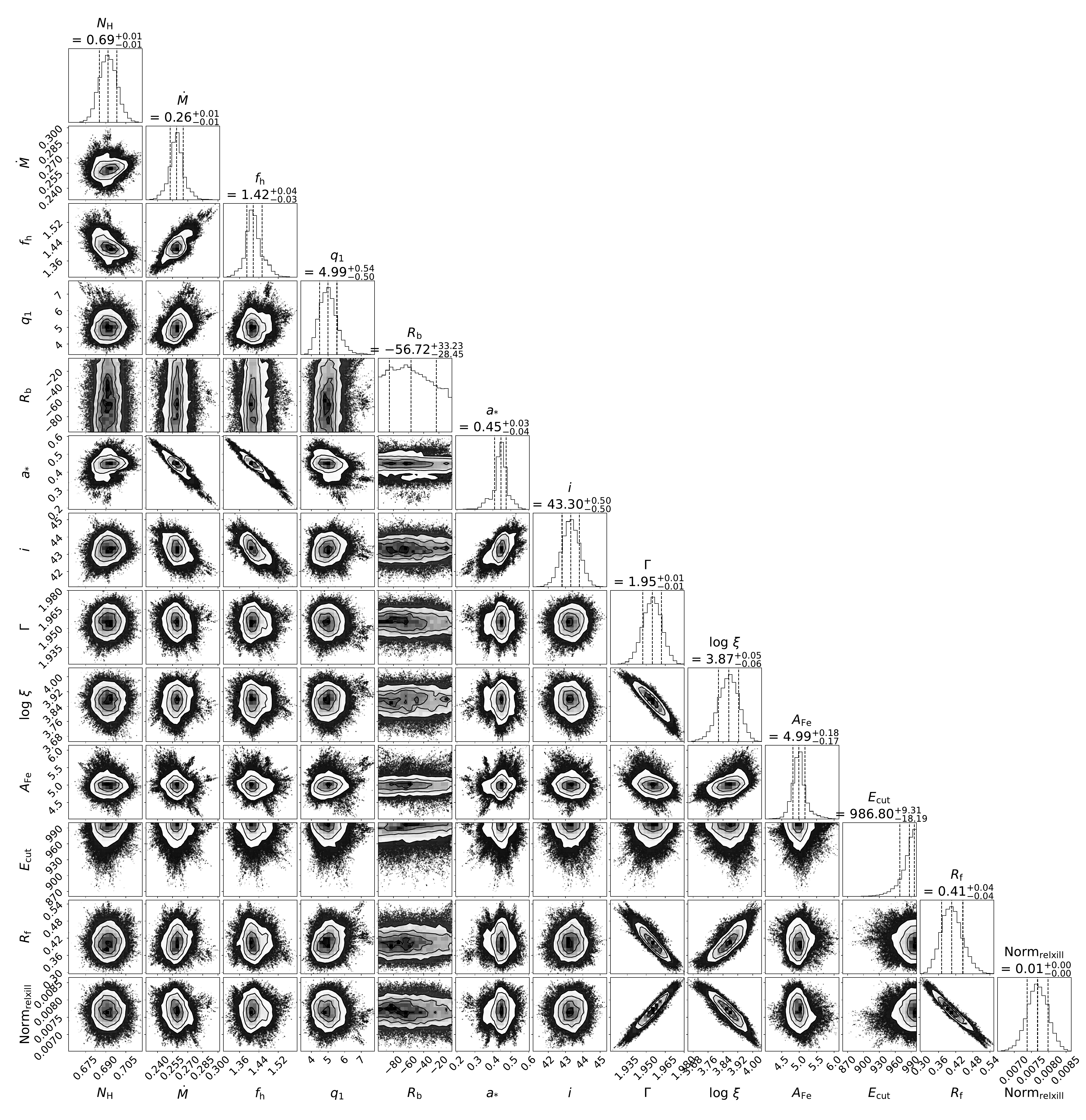}
    \caption{Probability distributions of the parameters for {\tt M1B} obtained from Data Set 1 through MCMC.}
    \label{fig:corner1}
\end{figure*}

\begin{figure*}
    \centering
    \includegraphics[width=1\textwidth]{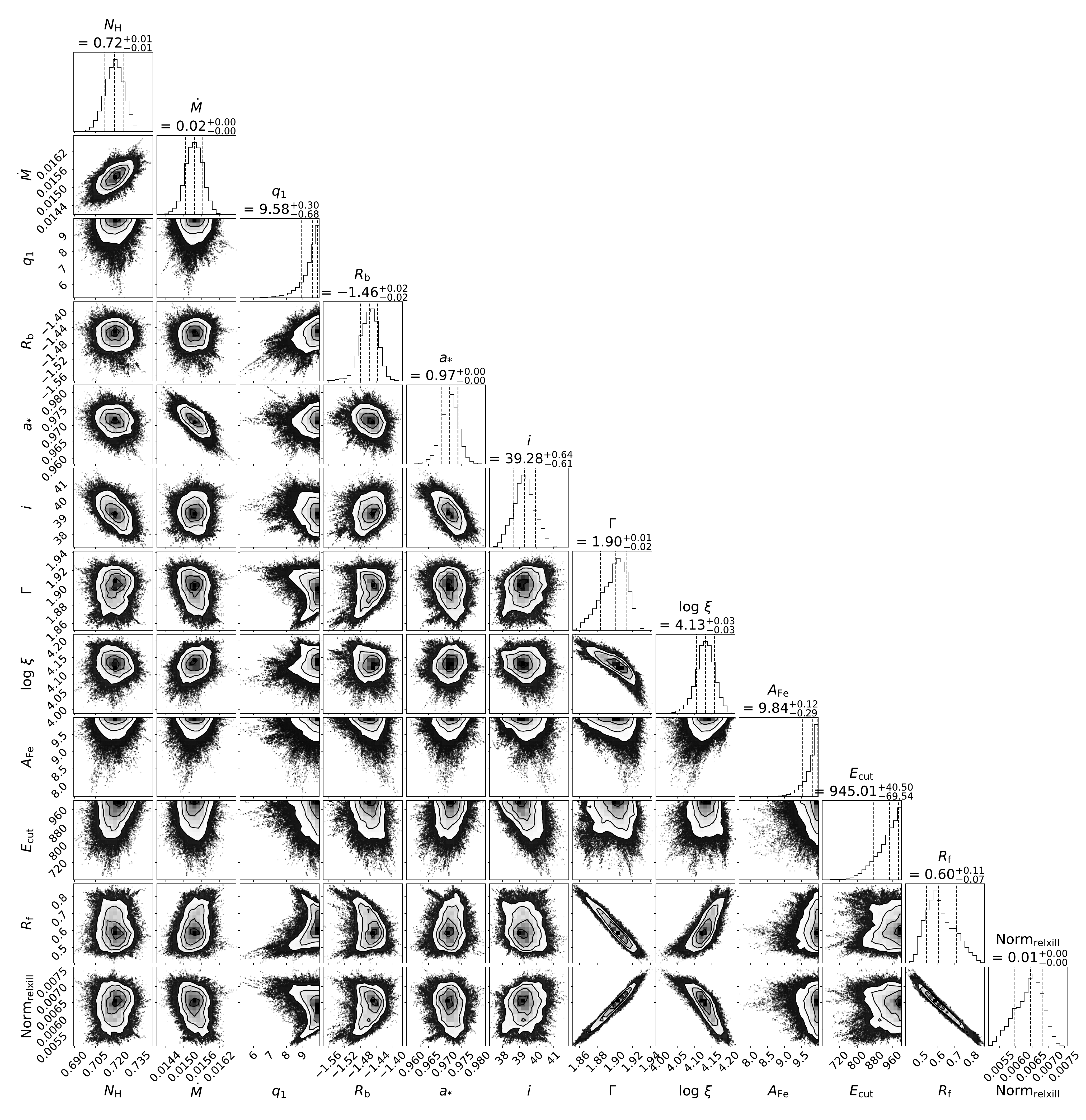}
    \caption{Probability distributions of the parameters for {\tt M2A} obtained from Data Set 1 through MCMC.}
    \label{fig:corner2}
\end{figure*}

\begin{figure*}
    \centering
    \includegraphics[width=1\textwidth]{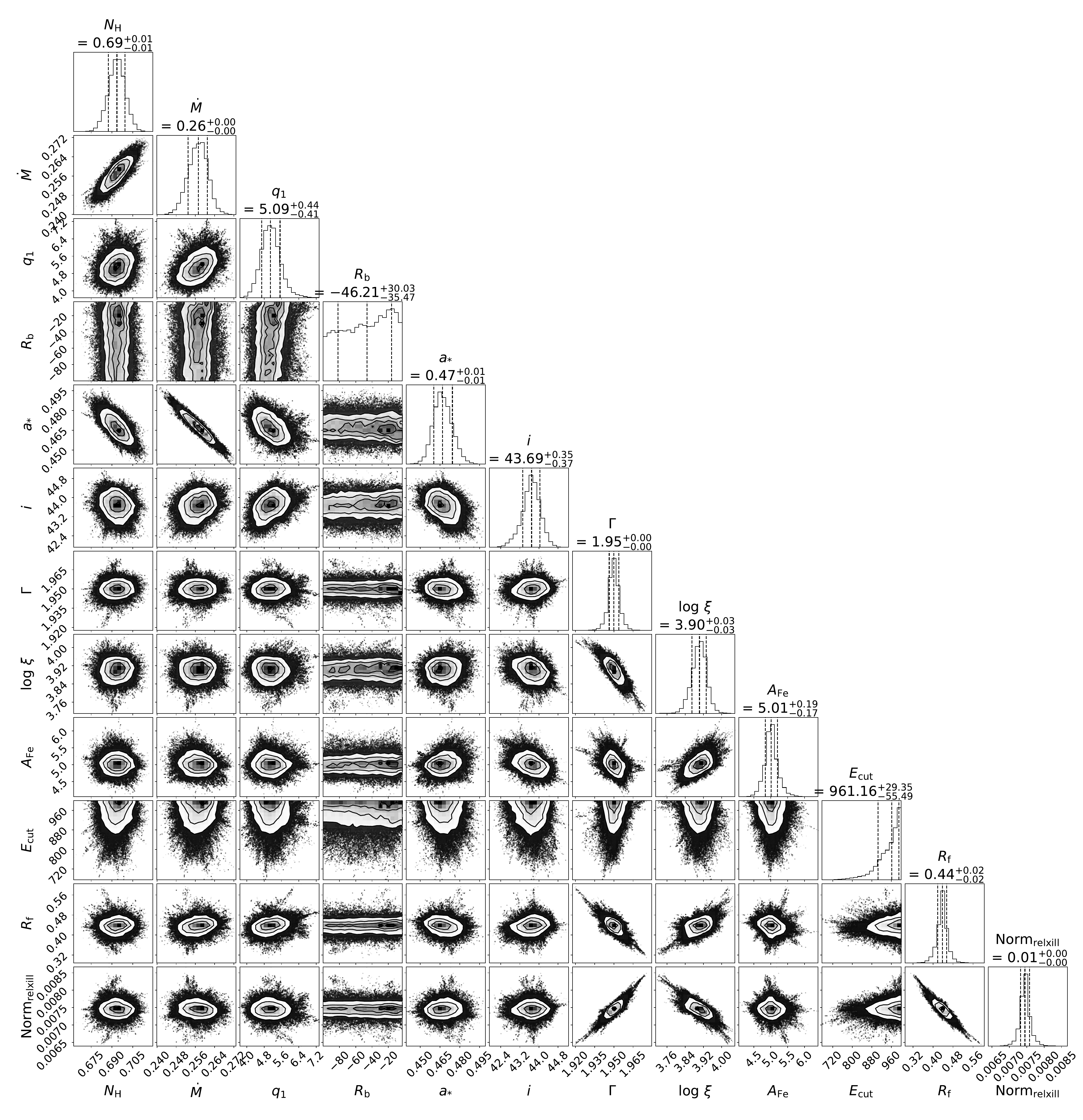}
    \caption{Probability distributions of the parameters for {\tt M2B} obtained from Data Set 1 through MCMC.}
    \label{fig:corner3}
\end{figure*}

\begin{figure*}
    \centering
    \includegraphics[width=1\textwidth]{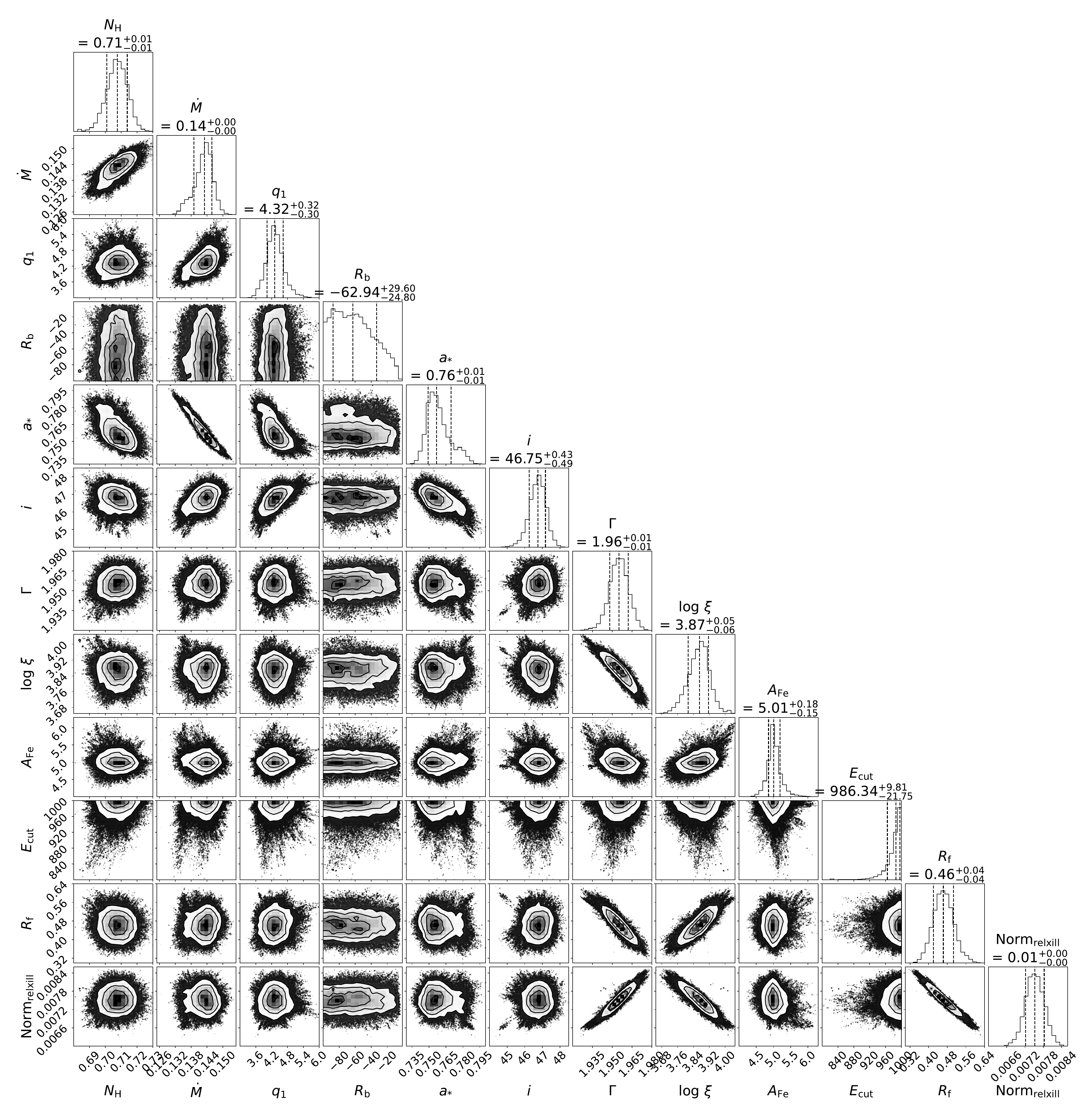}
    \caption{Probability distributions of the parameters for {\tt M3} obtained from Data Set 1 through MCMC.}
    \label{fig:corner4}
\end{figure*}
\end{appendix}
%-------------------------------------------------------------------
\end{document}